\documentstyle[12pt,aasms4]{article}

\begin{document}

\title{Classification of O Stars in the Yellow-Green: \\
        The Exciting Star VES~735} 

\author{C. R. Kerton\altaffilmark{1}, D. R. Ballantyne\altaffilmark{1}
and P. G. Martin}

\affil{Canadian Institute for Theoretical Astrophysics, University of
Toronto, Toronto, Ontario M5S~3H8, Canada}

\altaffiltext{1}{Also at Department of Astronomy, University of Toronto}

\authoremail{kerton, ballanty, pgmartin@cita.utoronto.ca}

\begin{abstract}

Acquiring data for spectral classification of heavily reddened stars
using traditional criteria in the blue-violet region of the spectrum can
be prohibitively time consuming using small to medium sized telescopes.
One such star is the Vatican Observatory emission-line star VES~735,
which we have found excites the H~II region KR~140.  In order to
classify VES~735, we have constructed an atlas of stellar spectra of O
stars in the yellow-green (4800 -- 5420 \AA).  We calibrate spectral
type versus the line ratio He~I $\lambda$4922:He~II $\lambda$5411,
showing that this ratio should be useful for the classification of
heavily reddened O stars associated with H~II regions.  

Application to VES~735 shows that the spectral type is O8.5.  The
absolute magnitude suggests luminosity class V.  Comparison of the rate
of emission of ionizing photons and the bolometric luminosity of
VES~735, inferred from radio and infrared measurements of the KR~140
region, to recent stellar models gives consistent evidence for a main
sequence star of mass 25~$M_\odot$ and age less than a few million years
with a covering factor 0.4 -- 0.5 by the nebular material.

Spectra taken in the red (6500 -- 6700~\AA) show that the stellar
H$\alpha$ emission is double-peaked about the systemic velocity and
slightly variable.  H$\beta$ is in absorption, so that the emission-line
classification is ``(e)''.  However, unlike the case of the more well
known O(e) star $\zeta$ Oph, the emission from VES~735 appears to be
long-lived rather than episodic.

\end{abstract}


\section{Introduction}
\label{sec:intro}

Accurate classification of OB stars associated with H~II regions is
important for the self-consistent modeling of the energetics.  As part
of our study of the isolated H~II region KR~140 (\cite{bal99}) we have
recently completed a series of spectroscopic observations to classify
its exciting star, VES~735 ($\alpha_{2000}=
2^{\mathrm{h}}20^{\mathrm{m}}5.37^{\mathrm{s}}$, $\delta_{2000}=
61^{\circ}7^{\prime}9.82^{\prime \prime}$). VES~735 was discovered and
catalogued in a Vatican Observatory objective prism emission-line survey
(\cite{coy83}).  Given its central association with an H~II region (see
Fig.~\ref{fig:radstar} and \S~\ref{sec:oestar}) and its emission-line
characteristics, VES~735 has to be an OBe star, but until now its
spectral classification has not been determined.

Due to the high level of reddening toward this star ($V = 12.88$, $B =
14.41$; \cite{bal97}) we were forced to find the temperature-related
classification using an alternative helium line ratio, He I
$\lambda$4922:He II $\lambda$5411 in the yellow-green spectral region.  In
\S~\ref{sec:atlas} we describe the atlas of spectra used to calibrate
the spectral type versus this helium line ratio. This calibration should
be useful for classification of faint O stars associated with H~II
regions; in \S~\ref{sec:class} we apply it to VES~735 in particular. In
\S~\ref{sec:oestar} we consider the emission-line properties of the
star.  Photometry of the star is used in \S~\ref{sec:photo} to obtain a
luminosity class for the star.  A detailed, multi-wavelength study of
the KR~140 H~II region and associated infrared nebula is presented in
another publication (\cite{bal99}).  In \S~\ref{sec:iflux} the radio and
infrared continuum measurements are summarized.  We discuss how these
might be used to investigate the rate of emission of ionizing photons
and bolometric luminosity of the exciting star, respectively, with
application to VES~735.  Conclusions are presented in \S~\ref{sec:conc}.
 
\section{Atlas of MK Standards in the Yellow-Green}
\label{sec:atlas}

Traditionally, MK classification of O stars is done in the blue-violet
spectral region (3800 -- 4900~\AA).  The ratio of the He I triplet line
$\lambda$4471 ($2^3P$ -- $4^3D$) and the He II $\lambda$4541 (Pickering
4 -- 9) line is used as the key temperature criterion (\cite{mor43};
\cite{con71}).  The heavy reddening of VES~735 made obtaining adequate
spectra (S/N$\sim$300 in the blue-violet) prohibitive on the 1.9-m
telescope at the David Dunlap Observatory (DDO; see Figure
\ref{fig:bluespec}), and so we decided to look for suitable helium line
ratios further to the red.

One possibility involves the lines He I $\lambda$5876 ($2^3P$ -- $3^3D$)
and He II $\lambda$5411 (Pickering 4 -- 7); this line ratio corresponds
very well with the spectral-type sequence (\cite{wal80}).  However, we
show in \S~\ref{sec:oestar} that VES~735 is an O(e) star with
double-peaked emission at H$\alpha$ throughout the time of our
observations.  Such stars can have He I $\lambda$5876 in emission
(\cite{con74}), and indeed we find that the corresponding singlet line
He I $\lambda$6678 line ($2^1P$ -- $3^1D$) is slightly in emission.
Therefore, since application of the calibration to VES~735 was our
primary goal, we decided to use the line He I $\lambda$4922 ($2^1P$ --
$4^1D$), the singlet line corresponding to the $\lambda$4471 triplet line.
Although this line is now slightly to the blue of $\lambda$5411, it is
still possible to obtain spectra of adequate S/N.

Due to the location of these lines between the blue-violet and the red,
there has been no work done on the systematics of the line ratio with
spectral type; therefore, we needed to calibrate the line ratio directly
using a series of stars of known spectral type.  Stars used in the
calibration of the helium line ratio are listed in Table~1
along with references for their MK classification. The stars observed
initially were all taken from Walborn's atlas of O star spectra
(\cite{wal90}).  Stars were then chosen from the other sources listed in
Table~1 to fill in gaps in spectral type coverage and to
provide duplication at some of the spectral types.  If there are two
different published spectral types for a given star, the spectral types
and sources are separated by a semicolon in Table~1. An O4
star, HD 46223, was also observed to illustrate the disappearance of He
I $\lambda$4922 by that spectral type.  Similarly, HD 22951 was
observed to show He II $\lambda$5411 disappearing by B0.5.

These stars were observed between October 1996 and December 1997 using a
Cassegrain-mounted CCD spectrograph at DDO (\cite{kam96}).  All of the spectra
have a dispersion of 32.5~\AA/mm and a resolution of $\Delta\lambda=2$~\AA.
The CCD data were reduced using standard IRAF procedures for the
processing of stellar spectra.  Spectra of
stars that were observed more than once, including VES~735, were
co-added after being brought to a common heliocentric radial velocity
scale.  The reduced spectra of the calibration stars are displayed
rectified and on a common scale in Figure~\ref{fig:atlas}.

\subsection{Calibration}
\label{sec:calib}

To provide a calibration of the spectral type versus the He~I
$\lambda$4922:He~II $\lambda$5411 line ratio, the equivalent widths ($W$)
of the $\lambda$5411 and $\lambda$4922 lines were first measured for all
of the calibration stars, using a line fitting program that fit a
Gaussian profile and a linear background simultaneously to the line and
the surrounding continuum.  Equivalent widths obtained this way were
compared with values obtained using an alternative technique where a
linear background was fit to the continuum and $W$ was simply the
integrated sum of the difference between the continuum fit and the data.
The two techniques agreed to within 1\% for all of the stars observed.
The W values and their statistical errors reported in Table~2
were obtained using the first technique.

In Figure \ref{fig:plotall} we show a plot of spectral type ($SpT$)
versus the logarithm of the line ratio $R = W_{4922}/W_{5411}$ for all
of the calibration stars.  In order to take into account uncertainties
in the spectral classification, the equivalent width ratio for those
stars with more than one published spectral type is plotted for each
spectral type. These stars are indicated by the joined data points. The
strong correlation ($r=0.965$) between $\log R$ and spectral type is
evident.  A least-squares fit was made to the data,
\begin{equation}
SpT = (9.04 \pm 0.10)+ (4.10 \pm 0.23) \log R,
\label{eq:allstars}
\end{equation}
and is shown in the figure as a solid line.  In the next subsection we
apply this calibration to our measurements of VES~735.  Note that the
choice of functional form does not matter; all that is required is a
good interpolation formula in the relevant range of the observed $R$.

Since the formal uncertainties in the W measurements are
small (typically 5\% or $\pm 0.02$ dex), most of the scatter we see in
the plots is intrinsic scatter.  Part of this scatter arises because different
classification schemes give slightly different results even for the same
star; also $R$ is not perfectly correlated with $SpT$ even for a
homogeneous set of classifications.  Thus, a given $R$ will correspond to
a small range of spectral types, as has already been noted for the He I
$\lambda$4471:He II $\lambda$4541 line ratio (\cite{con71}).  Finally,
$SpT$ for early stars is usually discretized into intervals of 0.5,
which will add to the scatter in any plot of $SpT$ against a smoothly
varying physical quantity.

\subsection{Spectral Classification of VES~735}
\label{sec:class}

VES~735 was observed in the yellow-green 32 times between January and
December 1997, resulting in a total integration time of about 19~h.  The
co-added spectrum of VES~735 is shown in Figure~\ref{fig:coadd}.

The vertical solid and dashed lines shown in Figure~\ref{fig:plotall}
represent our measurement of the line ratio $R$ and its standard error
for VES 735 (see Table~2).  Using Equation~\ref{eq:allstars}
we found $SpT = 8.3$.  Combining the uncertainty in our measurement of
$R$ for VES~735 and the standard errors from the calibration, we
estimate the total uncertainty in spectral type is $\pm 0.5$.  As noted,
in this temperature range spectral types are usually given with only 0.5
subdivisions; we conclude that VES~735 is an O8.5 star, with O8 and O9
being possibilities too.  Figure \ref{fig:comp} provides a comparison of
the VES 735 spectrum with those of the O7.5, O8, and O8.5 calibration
stars.  Since VES 735 is rapidly rotating (see \S\ \ref{sec:oestar}) the
comparison star spectra have been artificially broadened to simulate the
effect of rotation on the general appearance of the spectrum.

\section{VES~735 -- An O(e) Star}
\label{sec:oestar}

In addition to the yellow-green spectra obtained for classification
purposes we obtained H$\alpha$ spectra (6500 - 6700 \AA) of VES~735 in
October 1996, January 1997 and December 1997 at DDO.  The spectra were
obtained using the Cassegrain-mounted CCD spectrograph with a dispersion
of 10.2 \AA/mm and an instrumental resolution of $\Delta\lambda=0.5$
\AA.  Spectra at similar dispersion and a resolution of $0.7$ \AA\ were
also obtained in September 1997 at the Dominion Astrophysical
Observatory (DAO).  A series of red spectra (some co-added over several
days) of the star are shown in Figure \ref{fig:haspec} with notable
features highlighted.  A spectrum of the star BD+61 411 (spectral type
O8:) is shown for comparison.

The initial spectrum, in October 1996, was obtained to confirm the
emission-line character of the star as implied by its discovery in the
Vatican emission-line survey, and to demonstrate the association of the
star with the surrounding nebulosity as will be discussed.  The later
spectra from both DDO and DAO were obtained in order to monitor the
longevity and/or variability of the observed emission.

The most obvious feature visible in the red spectrum is the
double-peaked H$\alpha$ emission.  As in Be stars, this shape is
probably due to a rotating disk or shell of material around the star
(\cite{con74}). The two peaks are separated by about 400~km~s$^{-1}$ and
have shown variability in peak separation and height over the time
period of our observations (see Figure \ref{fig:haprof}).  From the
extent of the emission the rotating gas has $v \sin i \simeq 400$
km~s$^{-1}$, corresponding to an (inner) orbital distance of about $2
\times 10^{12}$~cm or about $4 R_{star}$.  Accounting for an underlying
absorption W value of about 2.6~\AA\ (estimated from BD+61~411)
the total W of the emission is 4.1~\AA. \footnote{In finding the
emission-line W values for Be stars, Dachs, Rohe, \& Loose
[1990], use absorption line W values from appropriately matched
stellar atmospheres rather than from comparison stars.}

Note also that H$\beta$ in VES~735 is much weaker in absorption than in
the calibration stars in Figure \ref{fig:comp}, and so is probably
partially filled in with emission.  The implied W value of the
H$\beta$ emission is about 1.5~\AA.

The January 1997 spectrum extends slightly redder, to include coverage
of entire He I $\lambda$6678 line to see if it would be useful for
classification purposes (see \cite{gro96} for digital spectra of dwarf O
stars in the red).  The spectrum shows that the He I $\lambda$6678 line
is being filled in by emission and therefore would not be useful.  The
implied W value of the emission is about 0.2~\AA.

Since we do not see any evidence for emission in the other lines of the
Balmer series, VES~735 is an O(e) star, using the notation of Conti and
Leep (1974) to denote visible emission in the Balmer series only at
H$\alpha$.  Oe and O(e) stars are the higher mass counterparts of the
more common and well-studied Be stars (\cite{fro76}). The double-peak
nature of the observed H$\alpha$ emission is similar to that observed in
the most well known example of an O(e) star, $\zeta$ Oph [O9~V(e);
\cite{nie74}].  Unlike VES~735, the H$\alpha$ emission from $\zeta$ Oph
has been very episodic, with emission outbursts lasting only a few
months and years passing between successive episodes (\cite{nie74};
\cite{ebb81}).  Incidentally, spectra taken of $\zeta$ Oph at DDO in
June 1997 showed no visible H$\alpha$ emission.

From the H$\alpha$ and H$\beta$ emission-line W values, and the
fluxes $F_\lambda$ of the stellar continuum at these wavelengths, the
Balmer decrement $D_{43} = (W F_\lambda)_\alpha/(W F_\lambda)_\beta =
0.9 \pm 0.1$; the uncertainty includes the measurement error and the
uncertainty in the effective temperature ($T_{eff}$) and the shape of
the stellar atmosphere flux distribution (relative $F_\lambda$).  This
flat (even inverted) decrement can be compared to values from a study of
Be stars by \cite{dac90}.  They find that the lowest decrements occur
for stars with relatively weak emission and for the early spectral
types; VES~735 appears to fit these trends, though it has a flatter
decrement than found in the Be stars (the lowest was 1.2).  Their
interpretation is that flat decrements like this arise due to optical
depth effects in circumstellar environments that are abnormally dense
and compact.  The circumstellar line-emitting plasma also has continuous
emission, most readily detected in the near infrared.  Their
interpretation is supported by the measured low ratio of line to
continuous emission in Be stars with a low decrement.

Note that if the optical light is contaminated by the continuous
emission of the circumstellar gas (perhaps 10 -- 20\% at $V$), then the
measured light is brighter and redder than the photosphere alone.  Both
effects would make the measured $M_V$ too bright (negative), by perhaps
as much as 0.3 mag.

The most common theory to explain the Oe and Be star phenomenon has to do
with mass loss from young stars.  Most Oe and Be stars are rotating very
rapidly and undergo periodic mass loss by some unknown mechanism (e.g.,
\cite{kam93}; \cite{how93}).  A simple measurement of the $\lambda$4922
line, assuming that it is purely rotationally broadened, gives $v \sin i
\sim 300$ km s$^{-1}$ for VES~735.  Material ejected from the rapidly
rotating star then orbits the star in a disk causing the observed
emission-line characteristics.  Winds from the star can eventually
disperse the orbiting material and cause the emission features to fade
and disappear (\cite{ebb81}).  Further monitoring of this interesting
star in the H$\alpha$ region would be of interest to see if the
apparently long-lived emission persists.

An alternative to consider is that VES~735 might be a higher mass
equivalent of a Herbig Be star, a very young star with a remnant disk of
material from its formation (\cite{wal97}).  However, the likely
evolutionary age of $3\times10^{5}$~y for the KR~140 H~II region argues 
against the star being extremely young (\cite{bal99}).

\subsection{Effect on our Spectral Classification}

Since we know that VES~735 is an emission-line star, of more concern to
us are any possible systematic errors in the measurement of $W_{4922}$
caused by the filling in of the He~I absorption line.  If there is such
contamination, then the line will appear to be too weak compared to the
He~II $\lambda$5411 line for which we do not expect any filling in, and
we will determine too early a spectral type for the star.  To
investigate this issue we plotted separately spectral type versus
$W_{4922}$ and $W_{5411}$ (see Figs.~\ref{fig:plotw4922} and
\ref{fig:plotw5411}).  Linear trends are obvious ($r=0.95$ and $r=-0.92$
respectively), and in both cases least-squares fits to the data were
made:
\begin{equation}
SpT = (4.82 \pm 0.03) + (7.92 \pm 0.56) W_{4922}  \label{eq:w4922},
\end{equation}
\begin{equation}
SpT = (12.80 \pm 0.41) - (7.08 \pm 0.58) W_{5411} \label{eq:w5411}.
\end{equation}
The vertical solid and dashed lines represent the measured values of $W$
for VES~735 and the standard error.  Using equations (\ref{eq:w4922})
and (\ref{eq:w5411}) we determine spectral types for VES~735 of O8.5 and
O7.5, respectively, each $\pm 1$ spectral type; note that the type based
on He~I $\lambda$4922 is later if anything.  Both are consistent with
the spectral type determined using equation (\ref{eq:allstars}).  We
conclude that filling in of the He~I $\lambda$4922 line is not a
significant factor in our spectral classification of VES~735.

\section {VES~735 --- Luminosity Class}
\label{sec:photo}

\subsection{Distance and Absolute Magnitude}

During the October 1996 and January 1997 observing sessions at DDO we
also obtained red spectra of BD~+61~411 which excites the neighboring
optical nebula IC~1795.  Sharp nebular lines of H$\alpha$ detected on
this spectrum and that of VES~735 (the KR~140 nebulosity) were used to
determine the differential radial velocity of the two nebulae (the
nebular lines are not visible in the ``sky''-subtracted spectra
presented here).  The velocity of IC 1795 is known from both H$\alpha$
and radio recombination lines to be $-42 \pm 2.0$ km s$^{-1}$
(\cite{cou66}; \cite{geo76}).  From this, the velocity of the KR~140
ionized gas was determined to be $-46 \pm 2.1$ km s$^{-1}$.  This is the
same as the velocity of another nearby ($\sim1^{\circ}$) nebula IC 1805
that from photometric studies of its star cluster (\cite{geo76}) is
known to be at a distance $d = 2.3 \pm 0.3$~kpc; the error estimate is
based upon the scatter between various published distance measurements
(see \cite{lei88}).  We adopt this as the distance to KR~140.  A
differential velocity of $+2.0 \pm 2.2$ km s$^{-1}$ was measured between
the nebular lines and the center of the stellar H$\alpha$ emission of
VES~735, consistent at least with a close association of VES~735 with
KR~140.  Given also the central position of VES~735 in the nebula
(Fig.~\ref{fig:radstar}), we therefore assume the same distance.

Photometry of VES~735 in the B and V bands was obtained for us at DAO:
$V=12.88 \pm 0.01$, $(B-V) = 1.53 \pm 0.02$ (\cite{bal97}).  Using
$(B-V)_\circ=-0.31$ (the intrinsic colors of O and very early B stars do
not vary significantly, \cite{joh67}) and ratio of total to selective
extinction $R_V=3.1 \pm 0.1$ we obtain a value for the total $V$ band
extinction, $A_V=5.7 \pm 0.2$.  Applying the usual distance modulus
equation we find $M_V=-4.6 \pm 0.2$ with an additional $\pm 0.28$
magnitudes from the uncertainty in distance.  As noted in
\S~\ref{sec:oestar}, this might be as low as $M_V=-4.3$ if there is
contamination by circumstellar continuous emission.

In Table~3 we summarize values of $M_V$ for O8.5~V and
O8.5~III stars from various sources.  Based upon our measurement of
$M_V$, and given this spectral type, VES~735 is luminosity class V.

\subsection{Checks on the Extinction}
\label{subsec:av}

An H$\alpha$ image of KR~140 was obtained for us at the Observatoire du
Mont Megantic (Fig.~\ref{fig:mm}; \cite{jon97}).  This faint nebulosity
is barely visible on the digitized red POSS plate, consistent with a
threshold in emission measure $E$ of approximately 100~cm$^{-6}$~pc
(\cite{van96}).  We also have a 1420 MHz continuum observation of the
region at lower resolution ($1'$; Fig.~\ref{fig:radstar}; \cite{bal99}).
We rebinned our radio image of the H~II region to the pixel size of the
H$\alpha$ image ($1.6''$).  Then, after removing point sources from the
H$\alpha$ image, we used
\begin{equation}
A_V = \frac{A_{H\alpha}}{0.82} = \frac{2.5}{0.82
}\log\left(\frac{E_{1420}} {E_{H\alpha}}\right)
\label{eq:av}
\end{equation}
on a pixel by pixel basis to construct an extinction map for the region.
Contours of $A_V$ are shown in Figure~\ref{fig:mm}.  Depending upon the
background we chose to subtract from the radio continuum image we
obtained a value of $A_V=5.8 \pm 0.2$ near VES~735.

We were also able to use the strong diffuse interstellar bands (DIBs) at
4430~\AA\ and 6613~\AA\ as another independent check on the amount of
reddening and extinction toward VES~735.  Although a few other DIBs
were tentatively identified in our spectra of VES~735, from the catalog
of Jenniskens \& D\'{e}sert (1994), they could not be measured precisely
enough to be useful.

For the $\lambda4430$ DIB, a correlation exists between percentage
central depth, $A_{c}$, and $E_{B-V}$ (\cite{sno77}): $A_c = 7.02E_{B-V}
+ 2.74,$ with rms dispersion $\sigma=1.65$.  In the VES~735 blue
spectrum we measured $A_c=15$, so that $E_{B-V}$=1.75 and $A_V=5.4$,
using $R_V=3.1$.
For the $\lambda$6613 feature we measured $W_{6613}=0.40$; using the
average $W/E_{B-V} = 0.231$ reported by Jenniskens \& D\'{e}sert (1994),
this corresponds to $A_V=5.4$.  These results all show that the value
derived for $A_V$ via photometry is reasonable.

\section{VES~735 --- Ionizing Photons and Bolometric Luminosity}
\label{sec:iflux}

As described below, radio continuum observations of H~II regions can
provide a measurement of the total flux of ionizing photons,
$Q(H^{\circ})$, of the exciting star(s).  We show how, in theory, the
combination of the observables $Q(H^{\circ})$ and $M_V$ can provide an
independent confirmation of both the luminosity and the temperature
classification of an OB star. (If there were other undetected O stars
then the $Q(H^{\circ})$ measured will be related to the combined output
of the stars and the technique clearly cannot be used to classify the
visible star.)  Similarly, the infrared re-emission by dust and
polycyclic aromatic hydrocarbons (PAHs) constrains the total luminosity.
We illustrate some of the uncertainties encountered in an actual
application to VES~735.

\subsection{The radio flux --  $Q(H^{\circ})$ relation}
\label{subsec:radio}

$Q(H^{\circ})$ is related to the observed radio continuum flux (F$_\nu$)
by
\begin{equation}
f Q(H^{\circ}) = \frac{\alpha_B F_\nu
d^{2}}{j_\nu} \label{eq:qflux}, 
\end{equation}
where $\alpha_B$ is the case B recombination coefficient of H$^+$
($3.27\times 10^{-13}$~cm$^{-3}$~s$^{-1}$ at 7500 K; \cite{sto95}),
$j_\nu$ is the free-free emissivity for H$^+$
($3.45\times10^{-40}$~erg~cm$^{-3}$~s$^{-1}$~Hz$^{-1}$~sr$^{-1}$ at
1420~MHz and 7500~K; after \cite{lan80}) and $f$ is the result of
several correction factors to be described.  Our measured value of
$F_\nu(1420~{\rm MHz}) = 2.35 \pm 0.05$~Jy (\cite{bal99}) then yields $f
Q(H^{\circ}) = (1.1 \pm 0.2) \times 10^{48}$~s$^{-1}$, or $\log[f
Q(H^{\circ})] = 48.05 \pm 0.1$.

A fraction of the star's ionizing photons might escape from the H~II
region, making the observed $F_\nu$ smaller (without affecting $M_V$),
if the nebula is not ionization bounded in the radial direction and/or
the covering factor is less than unity.  This can be described by a
factor $f_{ci} \le 1$.

Dust in the nebula competes with the gas for ionizing photons, producing
a factor $f_{dust} \le 1$.  As described by \cite{bot98}, the
effectiveness of dust depends on the optical depth to ionizing photons
(hence column density and shape of the absorption curve) across the
ionized volume, which in turn depends on the ionization parameter
(basically the ratio of the ionizing photon flux to the electron
density).  We have used version 90.04 of the spectral synthesis code
CLOUDY (\cite{fer98}) to model the specific case of KR~140 illuminated
by VES~735, matching both the flux and spatial distribution of surface
brightness in the radio and infrared (\cite{bal99}).  We estimate that
$\log(f_{dust}) \sim -0.1$ for dust like that in the diffuse interstellar
medium, where the absorption in the far ultraviolet is much larger than
in the optical; for dust like that found in dark clouds, where the absorption
increase into the ultraviolet is less pronounced, the competition for
ionizing photons is less efficient and $\log(f_{dust}) \sim -0.03$.

Finally, a factor $f_{complex}$ is needed to account for the fact that a
nebula is more complex than assumed by this approximation based on pure
H at constant temperature.  He is the next most relevant element here.
For a relatively low temperature star like VES~735 and a relatively low
density nebula like KR~140, most of the ionizing photons expended on
ionizing He ultimately produce H ionization as well (\cite{ost89}), and
so the total ionization rate of H is not much affected (nor is the total
emission of H$\beta$).  However, He is singly ionized in part of the
volume, and where present, He$^+$ increases the effective free-free
emissivity (the effect over the whole nebula KR~140 might amount to a
factor 1.06 [0.025 dex]).  Finally, $\alpha_B$ and $j_\nu$ have
different temperature dependences and the temperature is not constant in
a nebula; but the effect on the ratio is probably a few percent or less.

In general is hard to decouple the contributions $f_{dust}$ and
$f_{complex}$ entirely as they depend on and affect nebular structure.
However, their joint effect $f_{model}$ can be estimated using model
calculations.  Our best estimate from models which reproduce the size of
the nebula and the radio and infrared flux is $\log(f_{model}) \sim -0.04
\pm 0.04$.

Hence, $\log[f_{ci} Q(H^{\circ})] = 48.09 \pm 0.11$.  Paired with the
observed $M_V$, this is plotted in Figure~\ref{fig:mvq}, which has axes
oriented like a Hertzsprung-Russell diagram.  The error bars on the
point (a diamond symbol) represent those not related to the distance to VES~735.  Since
distance enters into the calculation of both $M_V$ and $Q(H^{\circ})$,
we indicate the effect of either increasing or decreasing the distance
to the star by two arrows (to the upper left and lower right,
respectively).  The length of the arrows represents a distance change of
$\pm 300$~pc, the estimated error, which affects $\log[Q(H^{\circ})]$ by
about $\pm 0.11$~dex.

\subsection{The infrared flux --  $L_{bol}$ relation}
\label{subsec:infrared}

Dust in and surrounding an H~II region absorbs radiation, both the
starlight directly and the diffuse (largely line) emission.  Measuring
the integrated dust radiation in the infrared, $L_{ir}$, gives an
estimate of the total, $L_{bol}$, reduced by the fraction $f_{cd}$.  The
criteria for $f_{cd}$ approaching unity are that in the radial direction
through the region actually being measured the absorption optical depth
be significant, particularly in the ultraviolet where most of the energy
is, and that the covering factor be large.  Therefore, $f_{cd}$ is akin
to $f_{ci}$, though as alluded to in \S~\ref{subsec:radio} most of
the absorption by dust occurs outside the ionized volume; with material
with adequate radial extent, probably the case here, they are both
equivalent to the covering factor. From models for which $f_{cd} = 1$,
we find that $\log(L_{ir}/L_{bol}) \sim -0.1$ because a small fraction
of the nebular gaseous re-emission is at long enough wavelengths to be
hard to absorb fully.

From HiRes images made from IRAS data, supplemented by the modeling
described which allows us to account for the unobserved emission at
longer wavelengths, we estimate that $\log(f_{cd} L_{bol}/L_\odot) =
4.55 \pm 0.1$ (\cite{bal99}).  Paired with $\log[f_{ci} Q(H^{\circ})]$,
this is plotted in Figure~\ref{fig:lq}.  The line to the upper left has
tic marks at intervals corresponding to 0.1~dex in each quantity,
whether from an error in distance or from a common covering factor.

\subsection{Stellar predictions of $M_V$ and $L_{bol}$ vs.\
$\log[Q(H^{\circ})]$} 

In Figure \ref{fig:mvq} we have plotted predictions based on OB star
models from \cite{vac96}, \cite{ost74} (reproduced in \cite{ost89}), and
\cite{pan73}.  The O8.5 predictions for each are circled.  Even with
precise models (and a large covering factor) it is readily apparent that
either $M_V$ or $Q(H^{\circ})$ alone is insufficient to obtain a full
spectral classification; a cooler giant star can have the same $M_V$ or
$Q(H^{\circ})$ as a hotter main-sequence star.  For example, in the
\cite{vac96} model, $M_V ({\rm O5~V}) = M_V ({\rm B0.5~III})$, and in
the \cite{pan73} model, $\log[Q(H^{\circ})] ({\rm O8~V}) =
\log[Q(H^{\circ})] ({\rm O9.5~III})$.  But by using both $M_V$ and
$\log[Q(H^{\circ})]$ one should, {\it in principle}, be able to remove
this degeneracy and obtain an independent estimate of the spectral
classification.

However, there are clearly significant disagreements between models.
They arise from different calibrations of $T_{eff}$ versus
spectral type, different $M_V$ calibrations versus spectral type, and
different stellar atmosphere models used in predicting the continuum
(see \cite{vac96} for a discussion of these differences and the
intrinsic uncertainties involved).  Systematic differences lead to
displacement and stretching (contraction) of the loci in a diagram like
Figure \ref{fig:mvq}.
There is also cosmic scatter of both $T_{eff}$ ($\pm 1700$~K) and $M_V$
($\pm 0.67$ magnitudes) at a given spectral type (\cite{vac96}) to be
considered along with the uncertainty of the calibration (average
value).  At O8.5~V, an uncertainty of 1700~K would correspond to an
error in $\log[Q(H^{\circ})]$ of 0.17~dex.

Figure~\ref{fig:lq} has loci based on the predicted $L_{bol}$ from
\cite{vac96} and \cite{pan73}.  Also plotted are a few of the
evolutionary models, with wind blanked atmospheres, from \cite{sch97}.
Model B1 is a for a 25~$M_\odot$ star at age $2.8 \times 10^4$~y.  B2
follows the evolution on the main sequence to age $2.6 \times 10^6$~y,
while B3 corresponds to a cooler subgiant at age $4.8 \times 10^6$~y.

\subsection{Discussion}
\label{sec:disc}

The \cite{ost74} model in Figure \ref{fig:mvq} suggests a classification
O9 -- 9.5~V for VES~735.  However, our spectroscopic observations are
not consistent with such a late O spectral type; nor are these models
based on the most current calibrations and atmospheres.

For the observed $M_V$, the \cite{pan73} models place the star near
O7.5, which is perhaps too early.  More importantly, this has the
implication that KR~140 is far from being ionization bounded
($\log[f_{ci}] \sim -0.6$).  The main reason to question such a low
value of the covering factor is the ring-shaped appearance of the radio
nebula, which looks like the projection of a somewhat broken thick shell
(Fig.~\ref{fig:radstar}).  If instead one starts with the spectral type
O8.5~V, then $\log(f_{ci}) \simeq -0.36$ and the observed $M_V$ is a bit
bright, perhaps overestimated because of continuous emission from the
circumstellar plasma (\S~\ref{sec:oestar}).

For the observed $M_V$, the \cite{vac96} models place the star at
O8.5~V, but again with a very small covering factor ($\log[f_{ci}] \sim
-0.6$).  There are two reasons that the \cite{vac96} models might have
overestimated $\log[Q(H^{\circ})]$.  First, the simple linear form of
their calibration of $T_{eff}$ versus spectral type, intended to
describe the trend over the whole range of OB stars, appears to
overestimate the actual $T_{eff}$ for stars near O8.5~V by typically
1500~K.  This would bring the scale closer to that adopted by
\cite{pan73} and \cite{lei90}, whose values of $T_{eff}$ and
$\log[Q(H^{\circ})]$ near O8.5~V agree well.  If so, then
$\log[Q(H^{\circ})]$ should move 0.15~dex to the right in
Figure~\ref{fig:mvq}.  Second, more recent stellar atmosphere models
which include wind blanketing have significant changes in the ionizing
continuum at a given $T_{eff}$ (\cite{sch97}).  Specifically, near
O8.5~V, $\log[Q(H^{\circ})]$ is lowered by 0.1~dex, moving the
prediction even further to the right to 48.47 as indicated by the arrow
in Figure~\ref{fig:mvq}.  While this is closer to the observed
$\log[Q(H^{\circ})]$, it still seems that $f_{ci}$ is considerably less
than unity ($\log[f_{ci}] \simeq -0.38$).

In Figure~\ref{fig:lq} we find evidence that $f_{cd}$ is also low, with
a value consistent with $f_{ci}$.  Accounting for a covering factor of
0.4 -- 0.5 would move the observed data point up by 0.3 -- 0.4 dex to
near the O8.5~V model of \cite{pan73} and B1 and B2 models of
\cite{sch97}.  Matching the \cite{vac96} O8.5~V model would require an
even lower covering factor, which the geometrical evidence suggests is
unlikely.  However, if the \cite{vac96} model were modified as described
above, it would move closer to the B2 model.  In the context of the
\cite{sch97} models, and the deduced covering factor, VES~735 is a
25~$M_\odot$ main sequence O star with an age less than a few million
years.

\section {Conclusions}
\label{sec:conc}

Calibration of the He I $\lambda$4922:He II $\lambda$5411 line ratio
using stars of known spectral type has allowed us to classify the
heavily reddened star VES~735 as O8.5. The atlas of stellar spectra in
the yellow-green along with the helium line calibration presented in
this paper should be useful for other studies of highly reddened O
stars, when obtaining reasonable classification spectra in the blue is
prohibitive.

Photometry of VES~735 and spectra taken in the H$\alpha$ region provided
additional information on the luminosity and emission characteristics:
VES~735 is an O8.5 V(e) star.  This direct identification of the
spectral type is useful in our modeling of the energetics of the
surrounding H~II region KR~140 (\cite{bal99}).  Likewise, the modeling
assists in assessing the ionizing luminosity and bolometric luminosity
from the radio and infrared observations.  There is consistent evidence
from comparing these luminosities with the predictions of recent stellar
models that VES~735 is a main sequence star of mass 25~$M_\odot$ and age
less than a few million years with a covering factor 0.4 -- 0.5 by the
nebular material.  The H$\alpha$ emission of this star appears to be
quite long lived compared with the O(e) star $\zeta$~Oph and continued
monitoring of its emission would be worthwhile.

\acknowledgements

We thank T. Bolton and S. Mochnacki for advice concerning the use of the
CCD spectrograph at DDO and subsequent data reduction, J. Thompson of
DDO for assistance with the observations, and D. Balam for photometric
data on VES~735.  Thanks also to R. Garrison and N. Walborn for helpful
comments regarding the spectral classification of OB stars.  This
research was supported by the Natural Sciences and Engineering Research
Council of Canada.  DRB participated via the Physics Co-op program of
the University of Victoria.

\clearpage

\figcaption[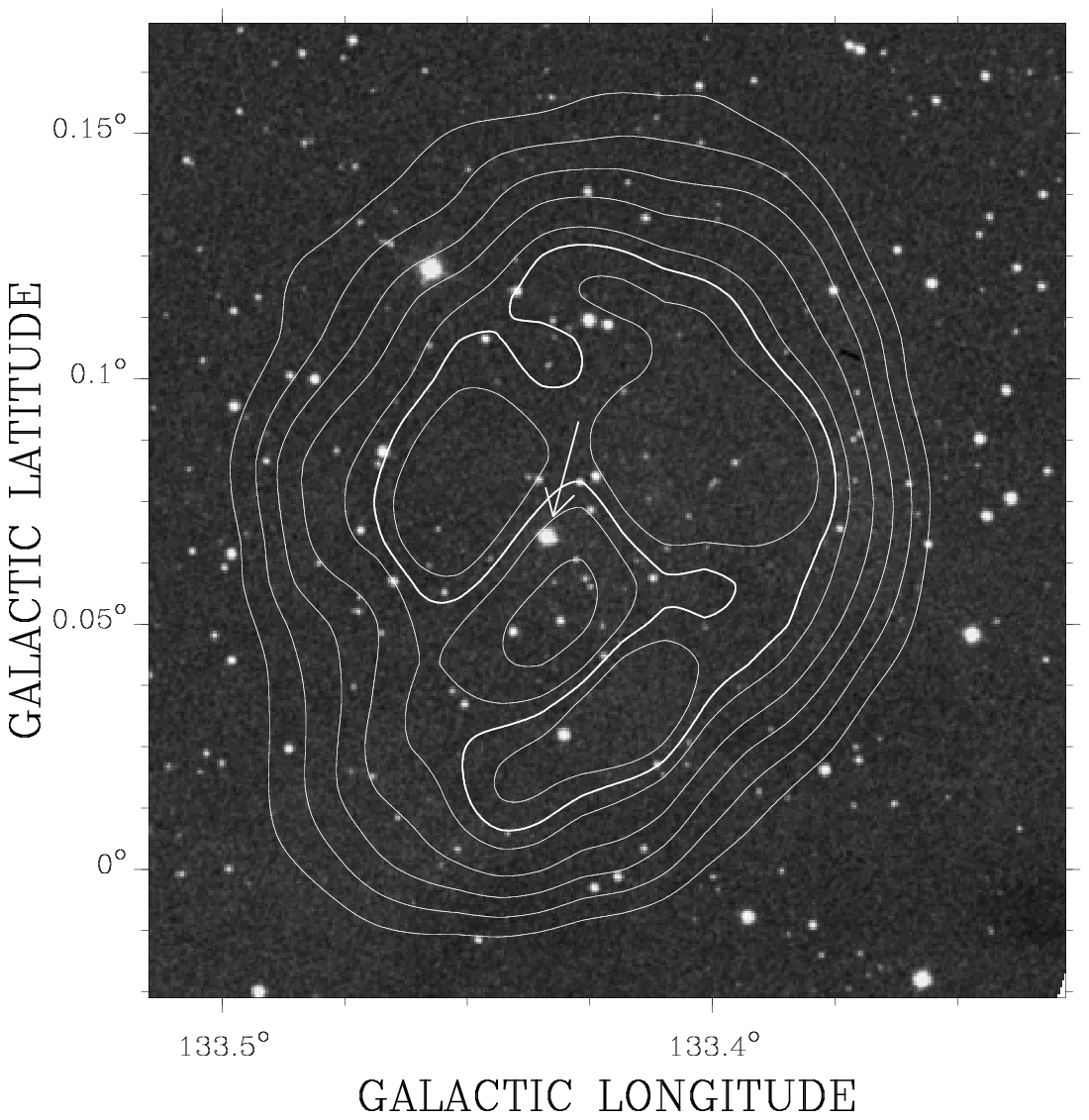]{Digital Sky Survey image of the KR~140 region
overlaid with contours of 1420~MHz continuum surface brightness
(observed with a $1^\prime$ beam), at 5, 6, 7, 8, 9, 9.5, and 10~K.
Note the ring-shaped appearance with central depression.  Location of
VES~735, the exciting star for this H~II region, is indicated by the
arrow.
\label{fig:radstar}}

\figcaption[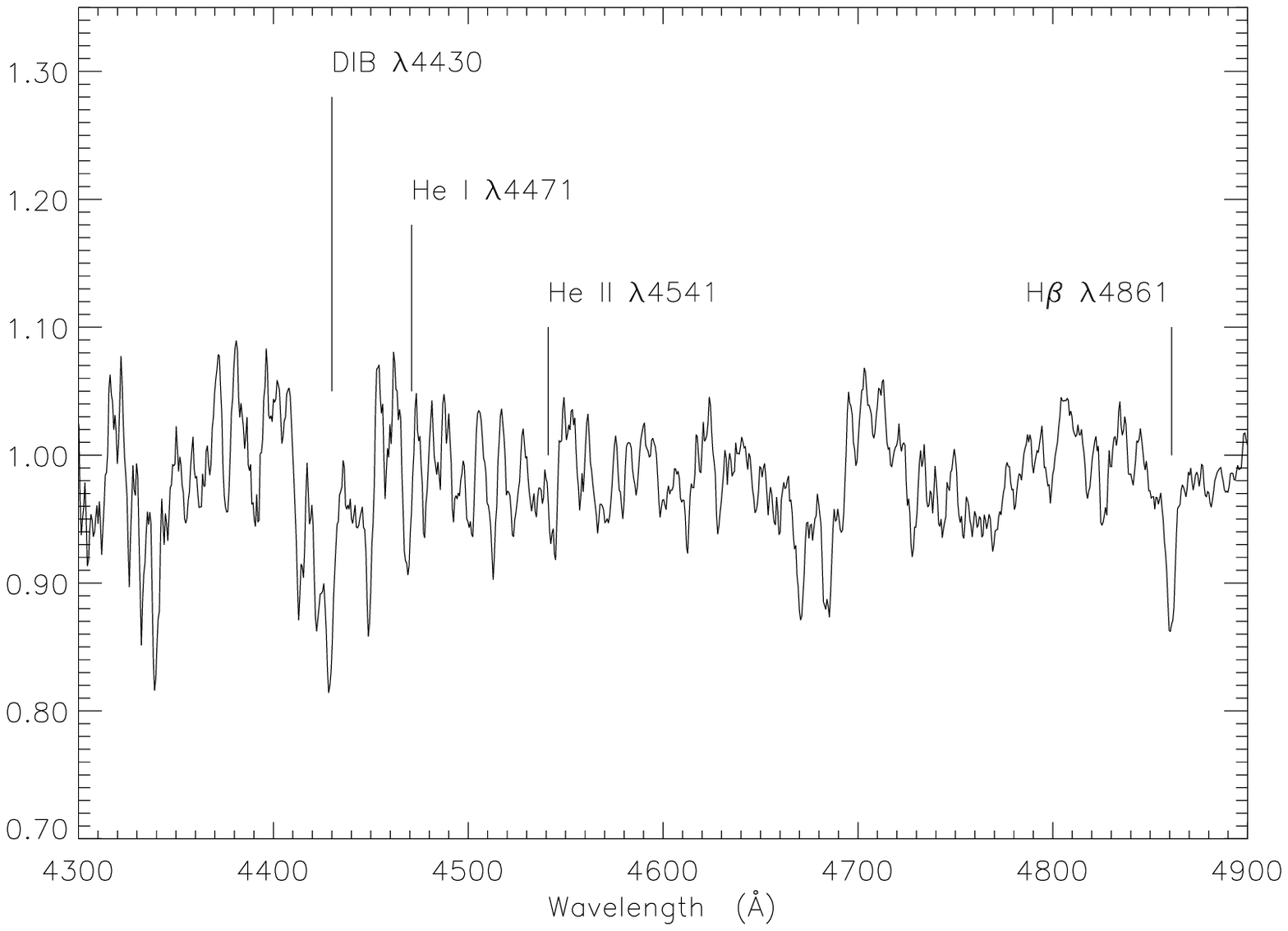]{Rectified, smoothed (boxcar = 9), co-added spectrum
of VES~735 in the blue spectral region (arbitrary flux units).  Normally
the He~I $\lambda$4471:He II $\lambda$4541 line ratio could be used for
spectral classification, but the star is so heavily reddened that even
this $\sim$7~h (co-added) exposure is not adequate.  The broad DIB at
$\lambda$4430 was used to estimate $A_V$ toward VES~735 (see
\S~\ref{subsec:av}).  \label{fig:bluespec}}

\figcaption[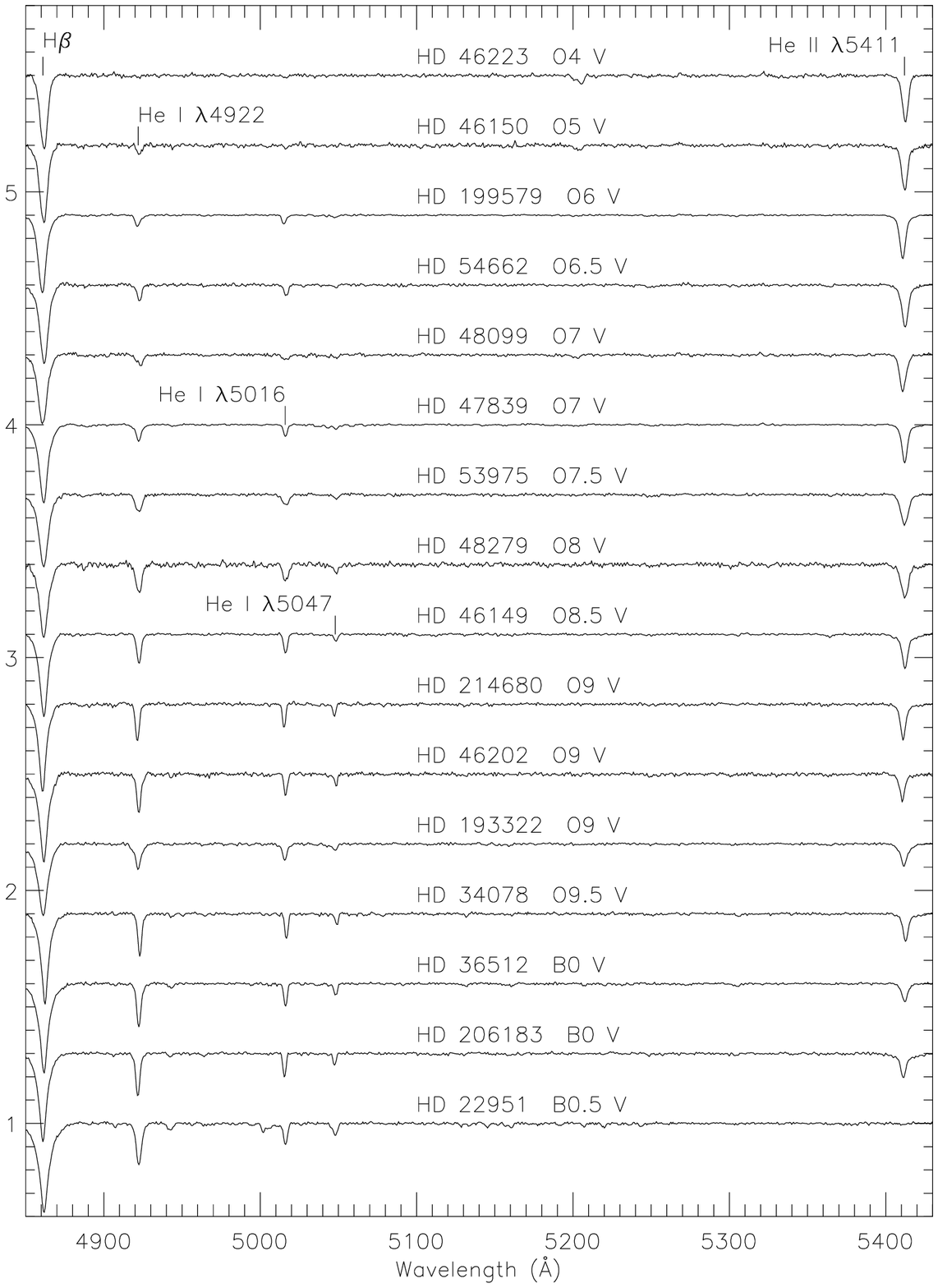]{Atlas of OB stellar spectra in the yellow-green.
Calibration stars used in this study are shown rectified and on a common
scale (with offsets of 0.3 units between spectra).  The O4 and B0.5
stars are included to illustrate the disappearance of the He~I
$\lambda$4922 and He~II $\lambda$5411 lines, respectively.
\label{fig:atlas}}

\figcaption[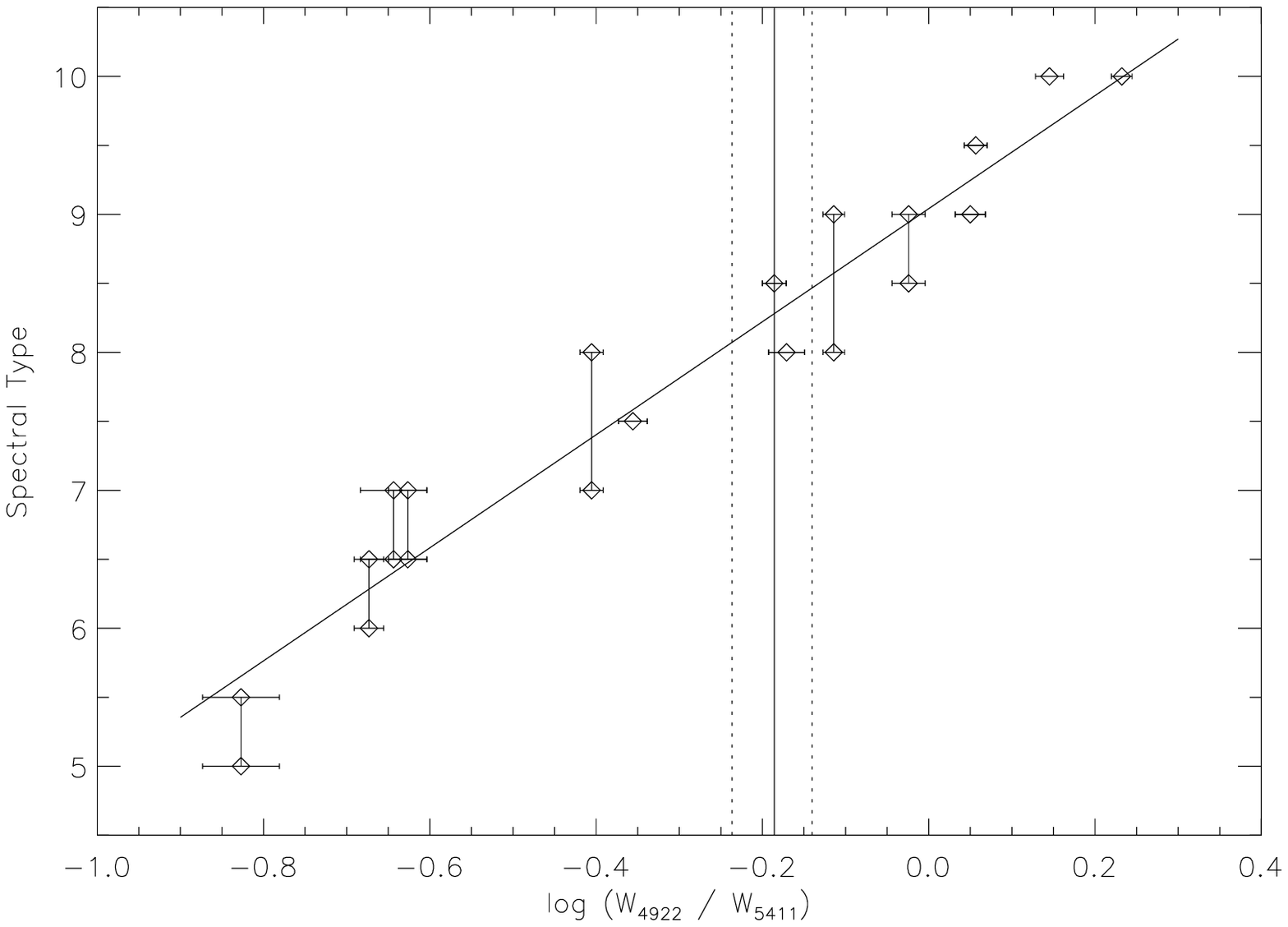]{Spectral type (O subtypes) vs.\ $\log$ of the 
ratio of the equivalent widths, $R = W_{4922}/W_{5411}$.  Spectral Type 10 
refers to B0 stars.  Stars with more than one  published spectral type are 
denoted by the joined data points.  The linear least-squares fit to the data 
is shown.  The vertical lines denote the measured $R$ with its standard error
for VES~735.
\label{fig:plotall}}

\figcaption[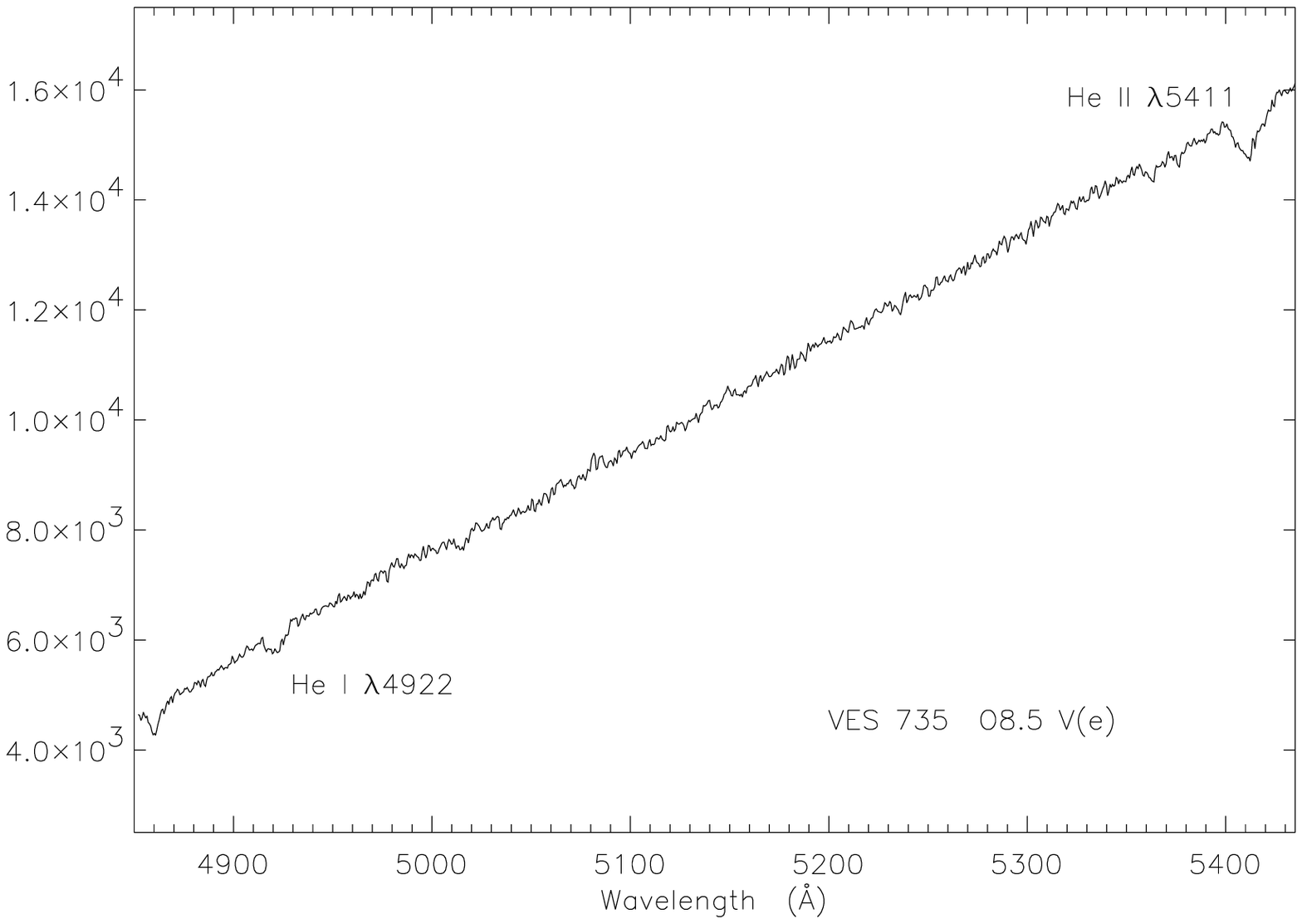]{Co-added ($\sim 19$~h) spectrum of
VES~735 in the yellow-green spectral region (arbitrary flux units).
\label{fig:coadd}}

\figcaption[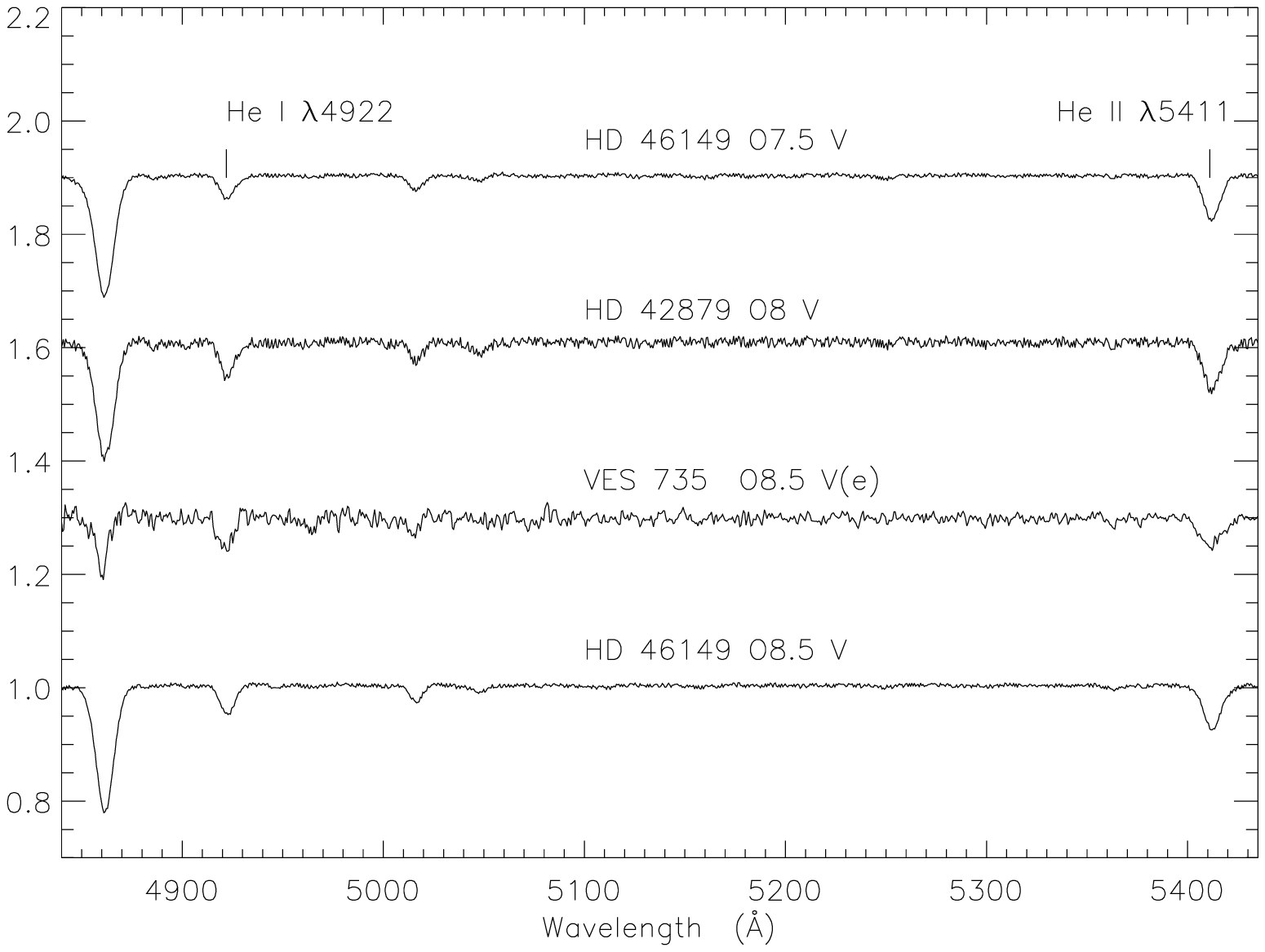]{Comparison of VES 735 spectrum with O7.5, O8, and
O8.5 stars (0.3 unit offsets). In order to simulate roughly the effect
of rotation present in VES~735, the calibration star spectra have been
convolved with a gaussian and artificial noise has been added to make
the S/N ratio equal to its original value.  \label{fig:comp}}.

\figcaption[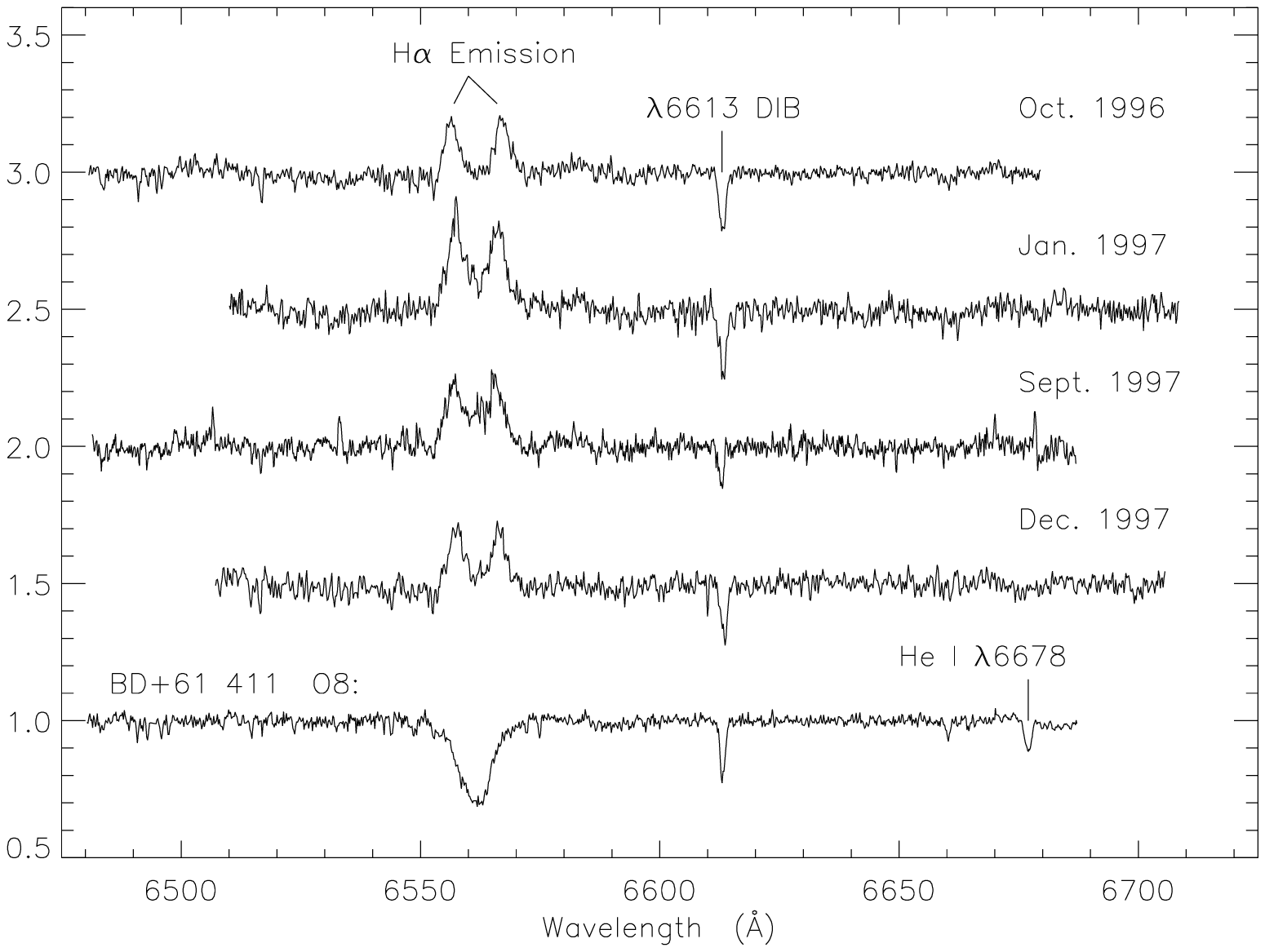]{H$\alpha$ spectra of VES~735 taken at DDO and DAO
(Sept.\ 1997 only).  The spectra are shown rectified and on a common
scale (with 0.5 unit offsets between spectra).  Note the long-lived,
strong, double-peaked H$\alpha$ emission.  The DIB at 6613~\AA\ was used
to provide another estimate of the amount of extinction toward VES~735
(see \S~\ref{subsec:av}).  For comparison, the O8: star BD+61 411 is
shown at the same scale; note the readily apparent He~I $\lambda$6678
line.
\label{fig:haspec}}

\figcaption[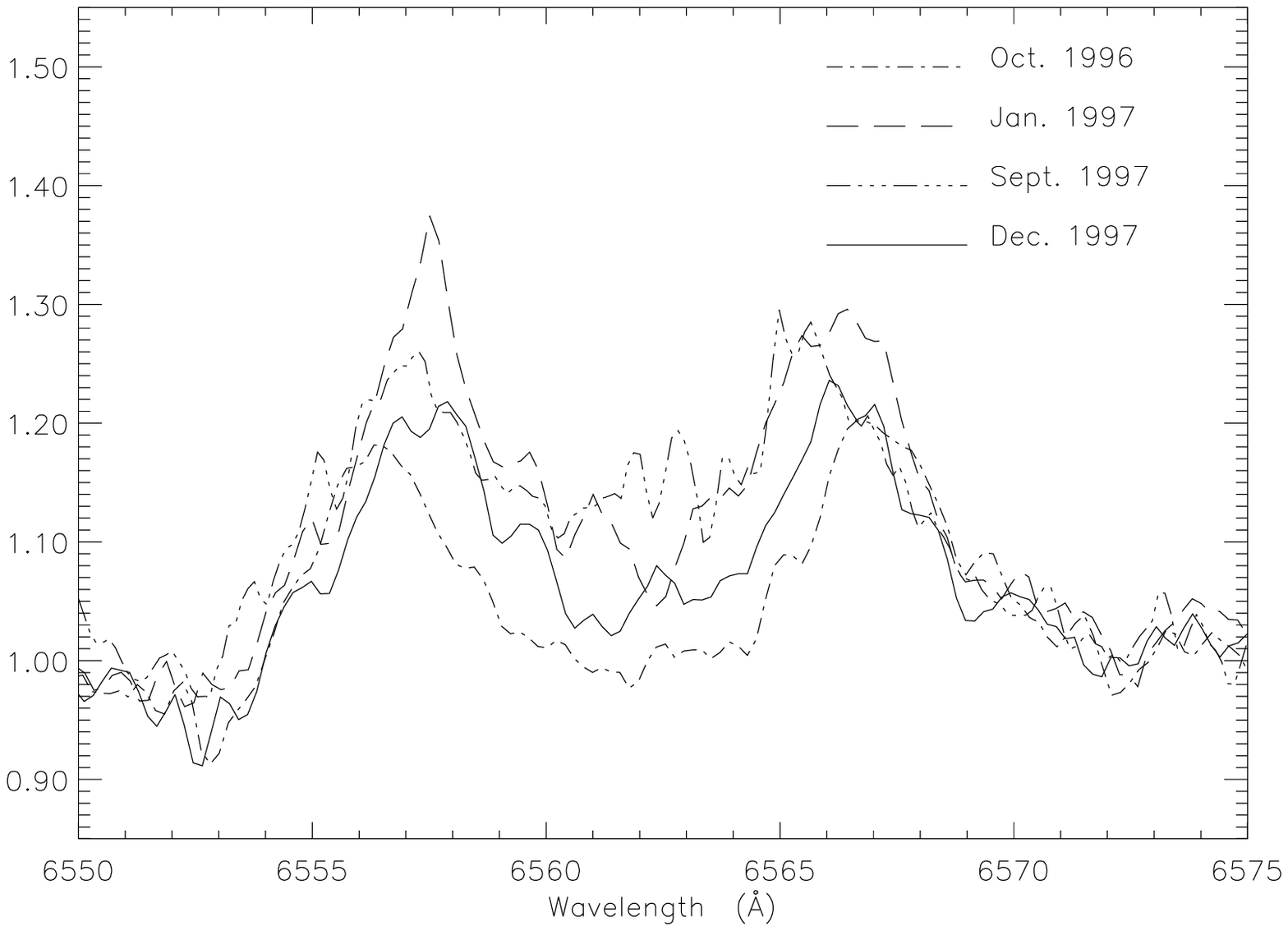]{Rectified VES~735 spectra highlighting the
H$\alpha$ region and illustrating the variations observed in the
line profile.  The spectra have been slightly boxcar smoothed
(boxcar = 3) for clarity.  \label{fig:haprof}}

\figcaption[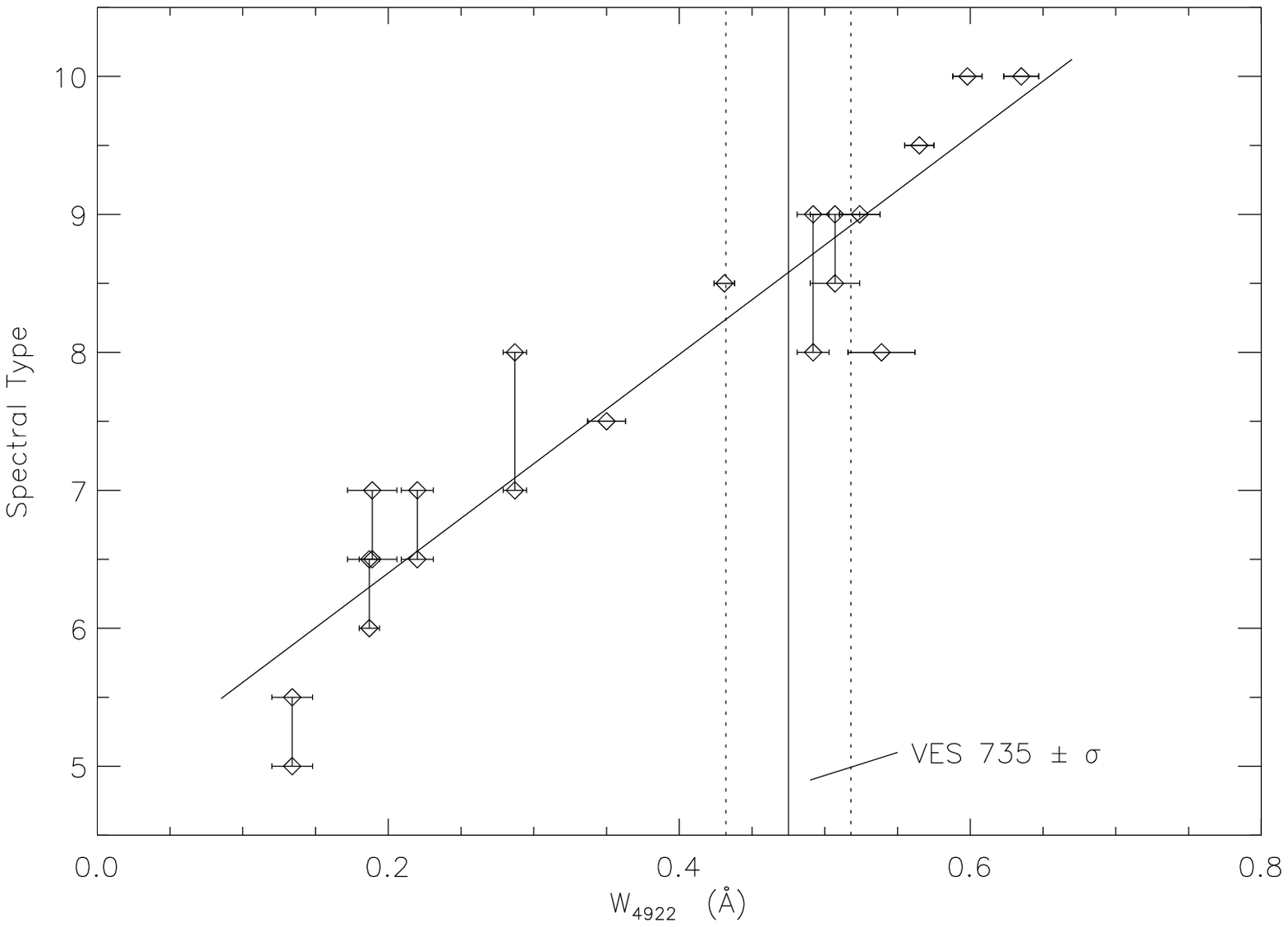]{Spectral type (O subtypes) vs.\ W$_{4922}$.  
Stars with more than one published spectral type are denoted by the joined
data points. Spectral Type 10 refers to B0 stars. A best fit line is shown.
The vertical lines denote the measured W$_{4922}$ and its standard error for 
VES~735.  \label{fig:plotw4922}}

\figcaption[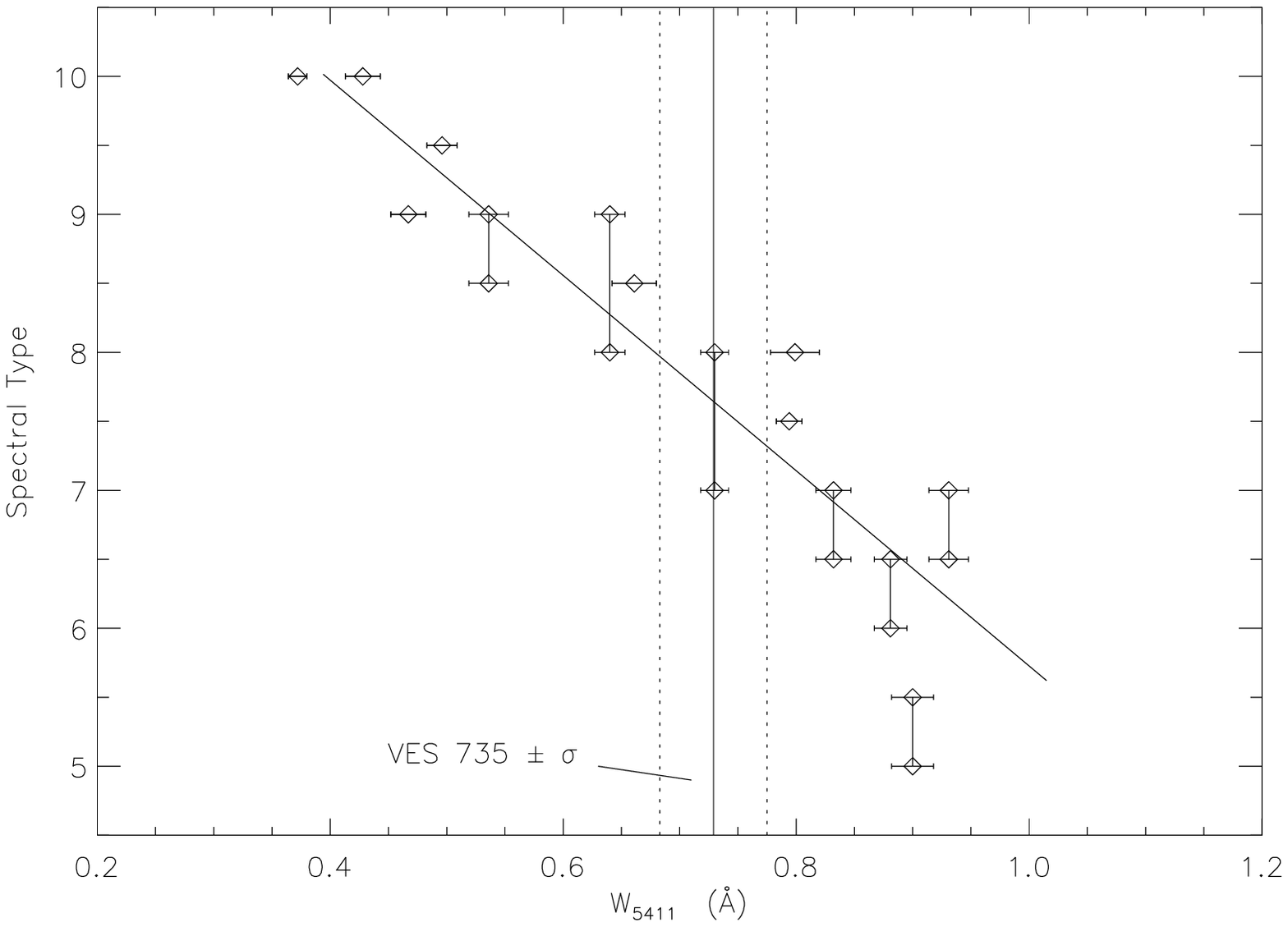]{Like Fig.~\ref{fig:plotw4922} but for
W$_{5411}$.  \label{fig:plotw5411}}

\figcaption[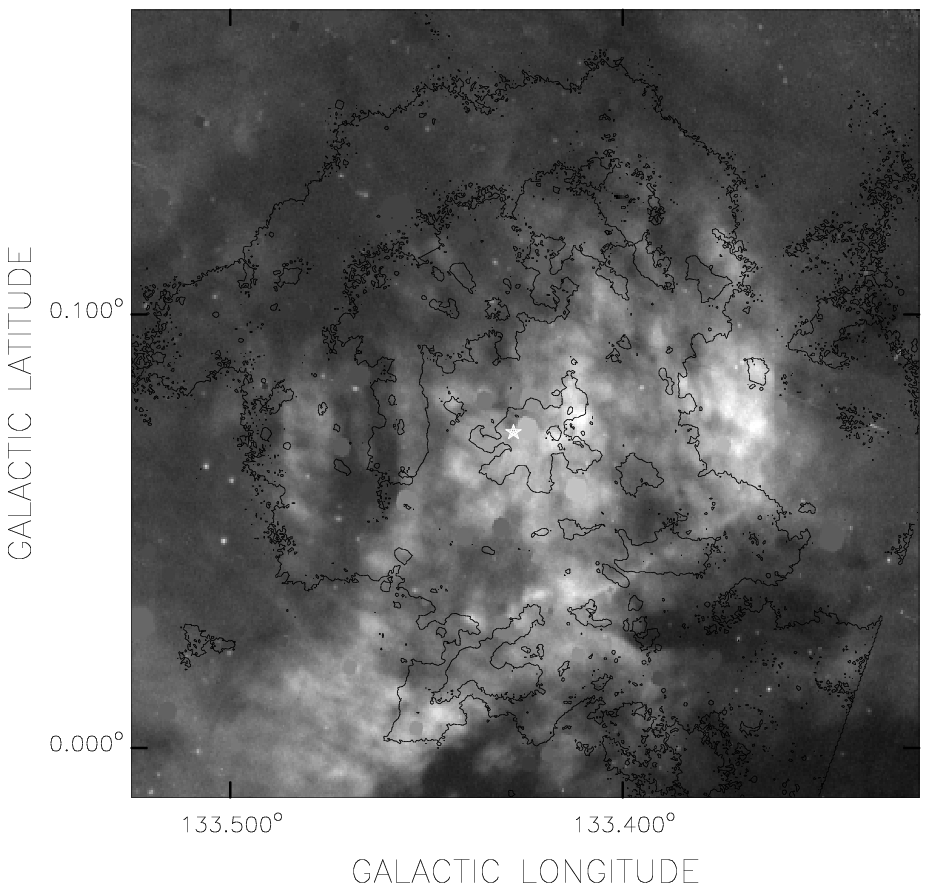]{H$\alpha$ image of the KR~140 region, taken by G. Joncas at
the Observatoire du Mont Megantic, used along with our 1420 MHz
continuum observations to put constraints on the amount of extinction
toward VES~735 (see \S~\ref{subsec:av}).  The brighter stars have been
removed from the image, and the position of VES~735 is indicated by the
star symbol.  Contours of total visual extinction are shown for $A_V$ =
6.0, 7.0, and 8.0.
\label{fig:mm}}

\figcaption[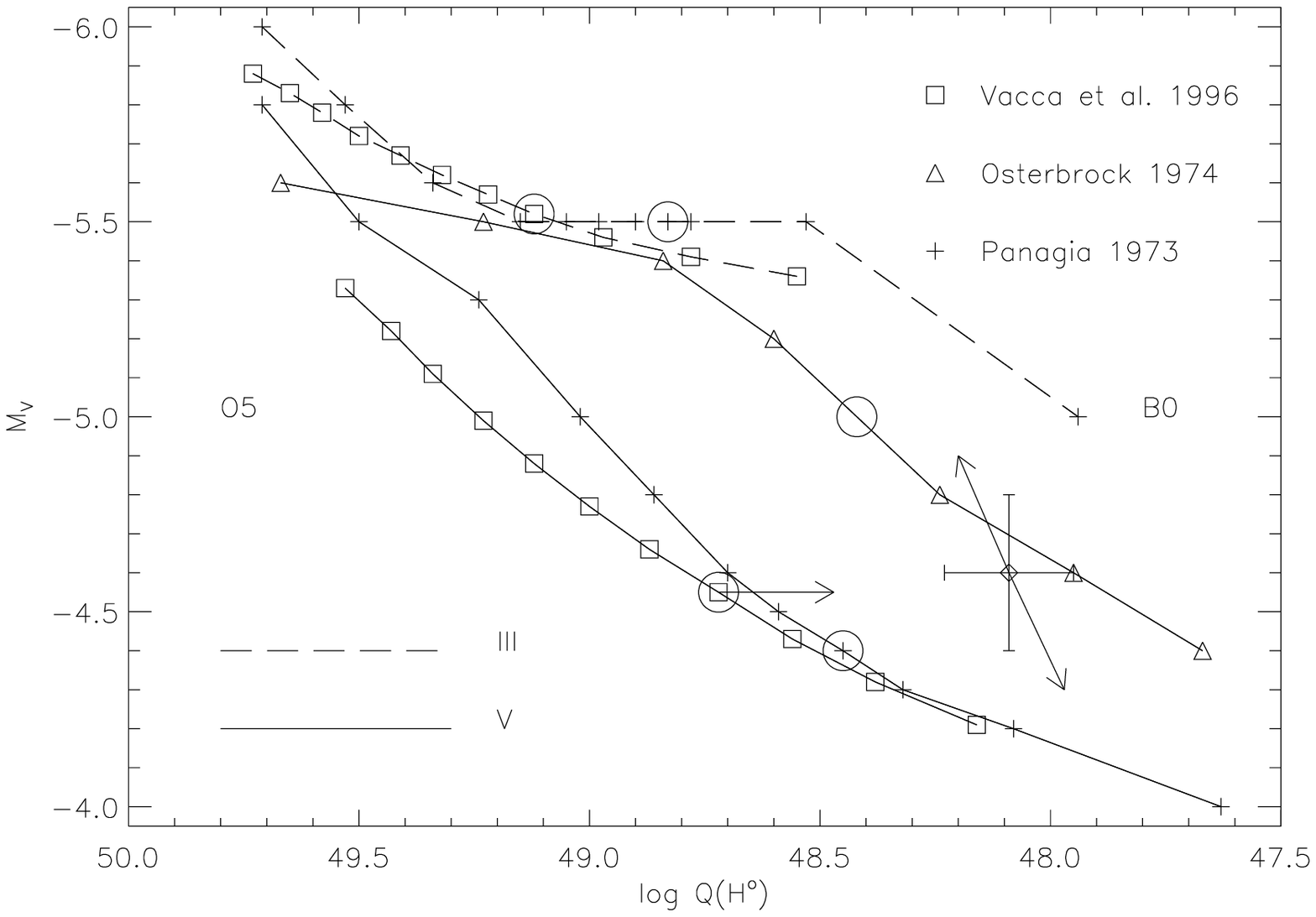]{$M_V$ vs.\ $\log[Q(H^{\circ})]$ from various
stellar model predictions.  Values deduced for VES~735 plotted as
a diamond, with error bars denoting errors not associated with the
distance; effect of increasing (decreasing) the distance by 300~pc
shown by the arrow to the upper left (lower right).  Models from
\cite{pan73} and \cite{vac96} are plotted at half-spectral-type
intervals from O5 to B0.  Models from \cite{ost74} are shown at
full-spectral-type intervals from O5 to O9, then at half-spectral-type
intervals.  The O8.5 models are circled.  The right arrow from the O8.5~V
model of \cite{vac96} estimates the combined effect of a lower effective
temperature calibration and using the wind-blanketed atmospheres of
\cite{sch97} (see \S~\ref{sec:disc}).  \label{fig:mvq}}

\figcaption[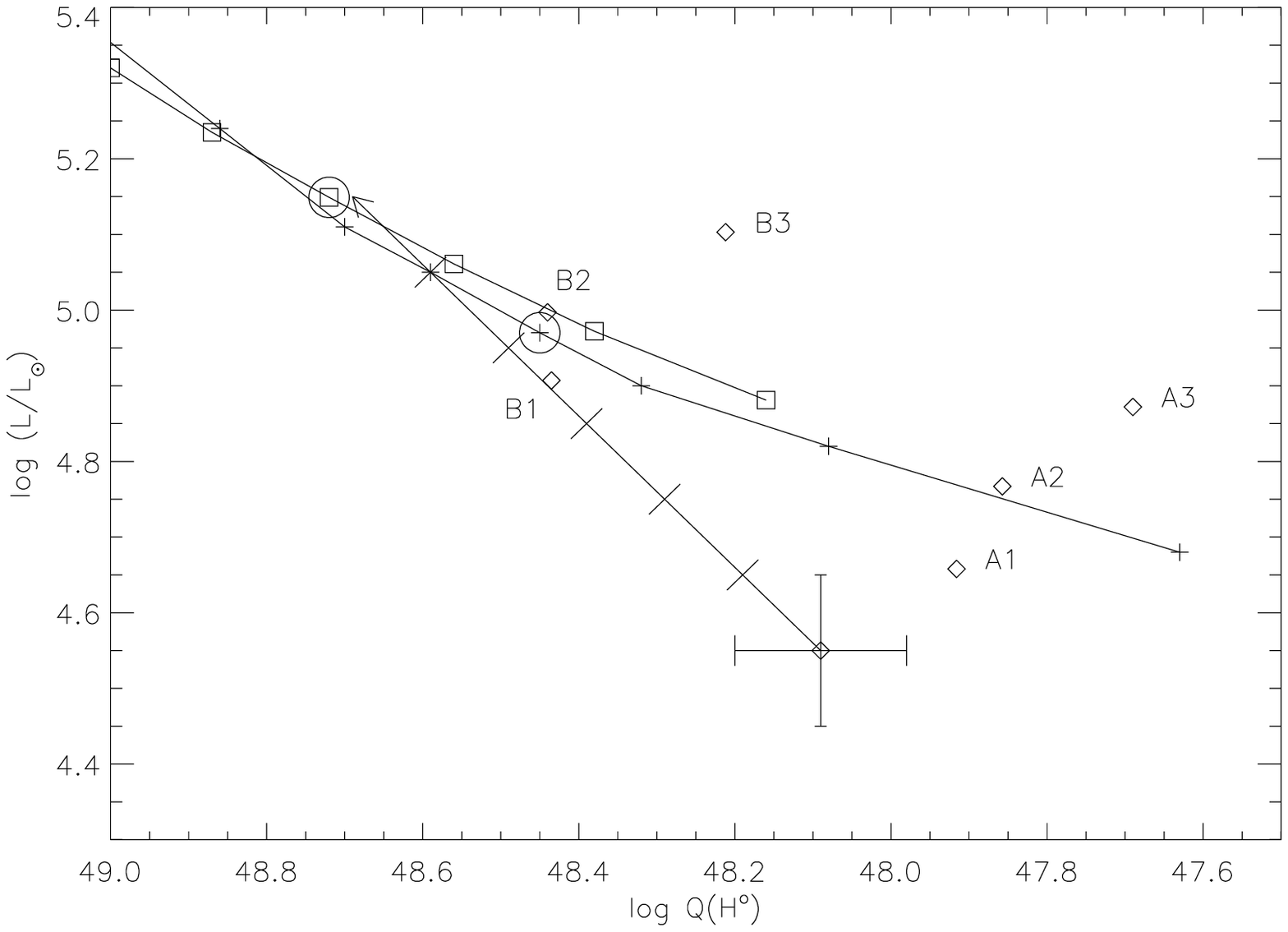]{Like Fig.~\ref{fig:mvq} but for
$\log(L_{bol}/L_\odot)$ vs.\ $\log[Q(H^{\circ})]$ and only main sequence
models from \cite{pan73} and \cite{vac96} supplemented by some
evolutionary models from \cite{sch97}.  The line to the upper left of
the diamond for VES~735 has tic marks corresponding to 0.1~dex
corrections to each ordinate that would result from common covering
factors $f_{ci}$ and $f_{cd}$.  \label{fig:lq}}


\clearpage

\begin{thebibliography}{}
\bibitem[Balam 1997]{bal97} Balam, D. 1997, private communication
\bibitem[Ballantyne et al.\ 1999]{bal99} Ballantyne, D. R., Martin,
P. G., \& Kerton, C. R. 1999, \apj, submitted  
\bibitem[Bottorff et al.\ 1998]{bot98} Bottorff, M., LaMothe, J.,
Momjian, E., Verner, E., Vinkovic, D., \& Ferland, G. 1998, \pasp,
110, 1040
\bibitem[Conti \& Alschuler 1971]{con71} Conti, P. S., \& Alschuler,
W. R. 1971, \apj, 170, 325  
\bibitem[Conti \& Leep 1974]{con74} Conti, P. S., \& Leep, E. M. 1974,
\apj, 193, 113 
\bibitem[Courtes et al.\ 1966]{cou66} Courtes, G., Cruvellier, P.,
Georgelin, Y., \& Astier, Y. 1966, J. Obs., 49, 329 
\bibitem[Coyne \& MacConnell 1983]{coy83} Coyne, G. V., \& MacConnell,
D. J. 1983, Vatican Obs. Publ., 2, 73 
\bibitem[Dachs et al.\ 1990]{dac90} Dachs, J., Rohe, D., \& Loose,
A. S. 1990, \aap, 238, 227
\bibitem[Ebbets 1981]{ebb81} Ebbets, D. 1981, \pasp, 93, 119
\bibitem[Ferland et al.\ 1998]{fer98} Ferland, G. J., Korista, K. T.,
Verner, D. A., Ferguson, J. W., Kingdon, J. B., \& Verner, E. W. 1998,
\pasp, 110, 761
\bibitem[Frost \& Conti 1978]{fro76} Frost, S. L., \& Conti,
P. S. 1976, Be and Shell Stars, IAU Symposium 70, A. Slettebak,
ed.\ (Dordrecht: D. Reidel) p.\ 139 
\bibitem[Georgelin \& Georgelin 1976]{geo76} Georgelin, Y. M., \&
Georgelin, Y. P. 1976, \aap, 38, 309 
\bibitem[Groppi \& Hansen 1996]{gro96} Groppi, C. E., \& Hanson, M. M. 1996,
\pasp, 108, 575
\bibitem[Howarth et al.\ 1993]{how93} Howarth, I. D. et al.\ 1993, \apj, 417, 338
\bibitem[Jenniskens \& D\'esert]{jen94} Jenniskens P., \& D\'esert,
F.-X. 1994, A\&AS, 106, 39 
\bibitem[Johnson 1967]{joh67} Johnson, H. L. 1967, \araa, 4, 193
\bibitem[Joncas 1997]{jon97} Joncas, G. 1997, private communication
\bibitem[Kambe et al.\ 1993]{kam93} Kambe, E., Ando, H., \& Hirata,
R. 1993, \aap, 273, 435 
\bibitem[Kamper 1996]{kam96}Kamper, K. 1996, DDO Observer's Manual, http://www.astro.utoronto.ca/$\sim$kamper/ddo\_man1.html
\bibitem[Lang 1980]{lan80} Lang, K. R. 1980, Astrophysical Formulae
(Berlin: Springer-Verlag) 
\bibitem[Leisawitz 1988]{lei88} Leisawitz, D. 1988, Catalog of Open
Clusters and Associated Interstellar Material (Greenbelt: NASA) 
\bibitem[Leitherer 1990]{lei90} Leitherer, C. 1990, \apjs, 73, 1
\bibitem[Morgan et al.\ 1943]{mor43} Morgan, W. W., Keenan, P. C., \&
Kellman, E. 1943, An Atlas of Stellar Spectra (Chicago: Univ. of Chicago
Press)
\bibitem[Niemel\"a \& M\'endez 1974]{nie74} Niemel\"a, V. S., \&
M\'endez, R. H. 1974, \apj, 187, 23L 
\bibitem[Osterbrock 1974]{ost74} Osterbrock, D. E. 1974, Astrophysics of
Gaseous Nebulae (San Francisco: W. H. Freeman)  
\bibitem[Osterbrock 1989]{ost89} Osterbrock, D. E. 1989, Astrophysics of
Gaseous Nebulae and Active Galactic Nuclei (Mill Valley: University
Science Books) 
\bibitem[Panagia 1973]{pan73} Panagia, N. 1973, \aj, 78, 929
\bibitem[Schaerer \& de Koter 1997]{sch97} Schaerer, D., \& de Koter,
A. 1997, \aap, 322, 598
\bibitem[Snow et al.\ 1977]{sno77} Snow, T. P., York, D. G., \& Welty,
D. E. 1977, \aj, 82, 113 
\bibitem[Storey \& Hummer 1995]{sto95} Storey, P. J., \& Hummer,
D. G. 1995, \mnras, 272, 41 
\bibitem[Vacca et al.\ 1996]{vac96} Vacca, W., Garmany, C., \& Shull,
J. 1996, \apj, 460, 914 
\bibitem[van Buren 1996]{van96} van Buren, D. 1996, private communication
\bibitem[Walborn 1980]{wal80} Walborn, N. 1980, \apjs, 44, 535
\bibitem[Walborn 1990]{wal90} Walborn, N. 1990, \pasp, 102, 379
\bibitem[Walborn 1997]{wal97} Walborn, N. 1997, private communication

\end{thebibliography}
\end{document}